\documentclass{iau} 
\usepackage{graphicx,txfonts,float,footmisc,natbib, url,booktabs}

\title[Hydrodynamic simulations of companion-perturbed AGB outflows] %% give here short title %%
{3D hydrodynamical survey of the impact of a companion on the morphology and dynamics of AGB outflows}

\author[J. Malfait \etal]   %% give here short author list %%
{Jolien Malfait$^1$, Silke Maes$^1$, Ward Homan$^2$, Jan Bolte$^1$, Lionel Siess$^2$, Frederik De Ceuster$^{3,1}$,
 \and Leen Decin$^{1,4}$}

\affiliation{$^1$ Institute of Astronomy, KU Leuven, \\ Celestijnenlaan 200D, 3001 Leuven, Belgium \\[\affilskip]
$^2$ Institut d'Astronomie et d'Astrophysique, Universit\'e Libre de Bruxelles (ULB), \\ CP 226, 1050 Brussels, Belgium \\[\affilskip]
$^3$ Department of Physics and Astronomy, University College London, \\ Gower Place, London, WC1E 6BT, United Kingdom \\[\affilskip]
$^4$ School of Chemistry, University of Leeds, \\ Leeds LS2 9JT, United Kingdom
\\ email: {\tt jolien.malfait@kuleuven.be}}

\pubyear{2021}
\volume{366}  %% insert here IAU Symposium No.
\jname{The origin of outflows in evolved stars}
\editors{L. Decin, A. Zijlstra \& C. Gielen, eds.}
\begin{document}

\maketitle

\begin{abstract}
With the use of high-resolution ALMA observations, complex structures that resemble those observed in post-AGB stars and planetary nebulae are detected in the circumstellar envelopes of low-mass evolved stars. These deviations from spherical symmetry are believed to be caused primarily by the interaction with a companion star or planet. With the use of three-dimensional hydrodynamic simulations, we study the impact of a binary companion on the wind morphology and dynamics of an AGB outflow. We classifiy the wind structures and morphology that form in these simulations with the use of a classification parameter, constructed with characteristic parameters of the binary configuration. Finally we conclude that the companion alters the wind expansion velocity through the slingshot mechanism, if it is massive enough. 
%% add here a maximum of 10 keywords, to be taken form the file <Keywords.txt>
\end{abstract}

\firstsection % if your document starts with a section,
              % remove some space above using this command.
\section{Introduction}
Low- to intermediate-mass stars shed their outer layers during the asymptotic giant branch (AGB) evolutionary phase through a dust-driven pulsation-enhanced wind \citep{Lamers1999,Hofner2018}. 
High-resolution observations reveal that these outflows contain a large diversity of complex structures, such as spirals, arcs, bipolarity, disks, etc. \citep{Ramstedt2014,Kervella2016,Decin2020,Homan2020a,Homan2020b}.
These observed AGB circumstellar envelopes (CSE) resemble the morphologies of planetary nebulae (PNe), and thereby help us fill the current knowledge gap about how the complex-structured planetary nebulae are shaped \citep{Sahai2011}.
The observed structures, that make the AGB wind deviate from spherical symmetry, are believed to be formed primarily by the interaction of the wind with a companion star or planet, that often remains undetected \citep{Decin2020}. 
A better understanding of how a companion can shape the winds of evolved stars is needed, since not accounting for the three-dimensional structures and the impact of a companion may lead to systematic errors in the estimate of critical stellar parameters such as the mass-loss rate.

Three-dimensional hydrodynamic simulations confirm that complex structures such as spirals and arcs form in stellar outflows when the impact of a companion is taken into account \citep{Theuns1993,Theuns 1996,Mastrodemos1998}. Depending on the wind characteristics and properties of the binary system, flattened or bipolar morphologies, and density enhancements around the orbital plane are predicted to form \citep{Mastrodemos1999,Kim2012,ElMellah2020}.
To improve our understanding on which binary and wind configurations create which type of wind structures and global morphologies, additional studies of high-resolution 3D hydrodynamic simulations are required. Here we discuss the main findings of such a study by \citet{Malfait2021} and \cite{Maes2021}, in which the wind structure formation of a set of simulations is studied in detail.

\section{Model grid}
The simulations are constructed with the three-dimensional smoothed-particle hydrodynamic (SPH) code \textsc{Phantom} \citep{Price2018}, which solves the fluid dynamic equations in a mesh-free way. The models are purely hydrodynamic, without the inclusion of dust, chemistry, radiation, and pulsations, and the cooling is regulated by the polytropic equation of state for an ideal gas, given by
\begin{equation}
\label{IdealGas}
P = (\gamma - 1) \rho u ,
\end{equation}
with polytropic index $\gamma=1.2$, and in which $P$ is the pressure, $\rho$ the gas density, and $u$ the specific internal energy. 
To improve these models and study the impact of dust, chemistry, radiation, pulsations and cooling, these missing ingredients are currently being implemented into \textsc{Phantom} by L. Siess, W. Homan and collaborators.
The simulations contain an AGB star with mass $M_{\rm{AGB}} = 1.5 \, \rm{M_\odot}$ that launches a wind of SPH gas particles. The grid of models consist of simulations characterised by a specific initial velocity $v_{\rm{ini}}$, companion mass $M_{\rm{comp}}$, orbital separation $a$, and orbital eccentricity $e$, as indicated in Table \ref{parameters}. The detailed setup of these simulations, together with an analyses of their wind morphology and dynamics, is described by \cite{Malfait2021} and \cite{Maes2021}.

\begin{table}[h!]
	\centering
	\caption{Characteristic model input parameters}\label{parameters}
	{\begin{tabular}{@{\extracolsep{\fill}}cccc}
			\midrule
			$v_{\rm{ini}} \, \rm{[km \, s^{-1}]}$&
			$M_{\rm{comp}} \, \rm{[M_\odot]}$& $a \, \rm{[au]}$&
			$e$\\
			\midrule
			 5&1   &2.5& 0.00\\
			10&0.01&4.0& 0.25\\
			20&    &6.0& 0.50\\
			  &    &9.0&     \\
			\midrule
	\end{tabular}}
%	\tabnote{\textit{Notes}: [1] 13. Commission des \’eclipses solaires [2] 13. Commission Eclipses of Sun}
\end{table}

\section{Morphology classification}
The companion shapes the AGB wind morphology by two primary effects, namely the induced orbital motion of both components around their center-of-mass (CoM), and the gravitational attraction of wind particles. The physical properties of the binary system and the AGB wind will determine the relative strength of these effects and the resulting global shape of the outflow.
In general, the perturbation by the companion is stronger when its mass $M_{\rm{comp}}$ is large, the orbital separation $a$ small, the AGB wind velocity $v_{\rm{ini}}$ low, and the eccentricity $e$ high. To estimate the degree of complexity induced by the companion, these parameters are combined in a classification parameter $\varepsilon$, which is defined as
\begin{equation}
\label{vareps}
\varepsilon = \frac{e_{{\rm grav}}}{e_{{\rm kin}}} = \frac{ \frac{G M_{\rm{comp}} \rho} {R_{\rm Hill}}}{\frac{1}{2} \rho v_{\rm w}^2} = \frac{(24 G^3 M_{\rm{comp}}^2 M_{\rm{AGB}})^{1/3}}{v_{\rm w}^2 a (1-e)},
\end{equation}
so the ratio of the gravitational energy density of the companion to the kinetic energy of the wind \citep{Maes2021}.
%\footnote{
%The wind velocity $v_{\rm w}$ is defined as $
%v_{\rm w} = \sqrt{v_{\rm single}^2(r = a) + v_{\rm p}^2}$,
%with $v_{\rm single}(r)$ the wind velocity at radius $r$ in an isotropic, single-star model with the same input wind velocity and AGB star as the binary model.}
A low $\varepsilon$ value corresponds to a limited impact of the companion on the wind dynamics and morphology, whereas high $\varepsilon$ values indicate that the wind will be strongly perturbed.

\subsection{Wind structure around companion}
The simulations are categorised according to the inner wind structure, that forms around the companion star and shapes the global wind morphology, which becomes more complex for increasing $\varepsilon$ value. Note that the exact $\varepsilon$ values delimiting these three categories are uncertain, and based on the available simulations of the studies by \citet{Malfait2021} and \citet{Maes2021}.

\begin{figure}[h!]
	\centering
	\includegraphics[width = \textwidth]{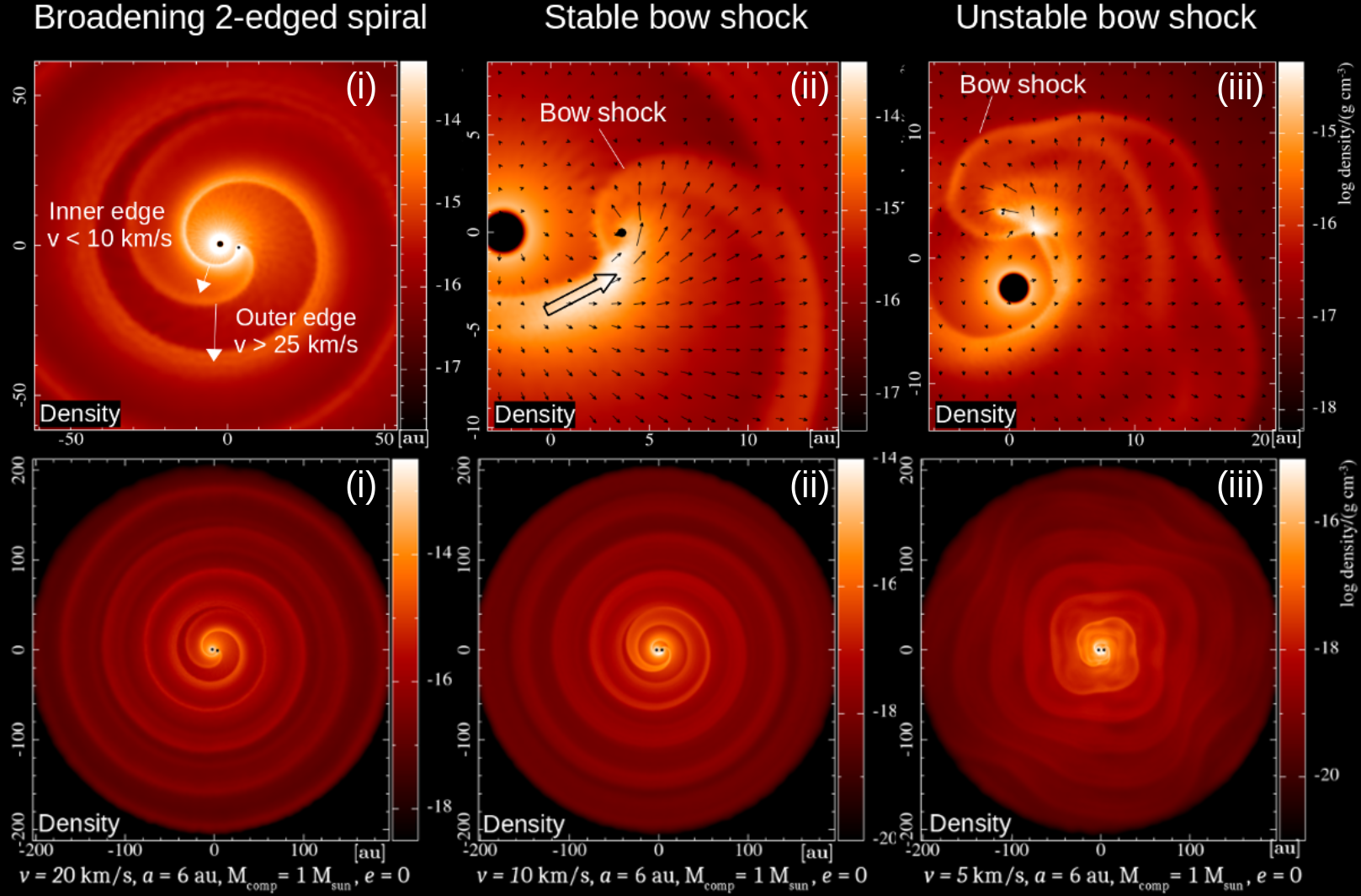}
	\caption{Density distribution in a slice through the orbital plane of three simulations with $a = 6 \, \rm{au}$, $M_{\rm{comp}} = 1 \rm{M_\odot}$, $e  = 0$ and from left to right $v_{\rm{ini}} = 20, 10, 5 \, \rm{km \, s^{-1}}$. The upper layer shows the inner density structures formed around the companion, the lower layer shows the global orbital plane morphology. Figure adapted from \citet{Malfait2021}.}
	\label{plotMorph}
\end{figure}

By studying the wind structures in a slice through the orbital plane of the 3D morphology, we find the following classification:
(i) For configurations with a classification parameter $\varepsilon \lesssim 1 $ (illustrated in the left column of Fig. \ref{plotMorph}), a broadening spiral structure forms attached to the companion, that is delimited by a slow, dense inner edge and a higher-velocity, less dense outer edge. This inner wind structure shapes the outflow into an approximate Archimedes global spiral structure.
(ii) In case of a stronger wind-companion interaction intensity, so higher $\varepsilon$-value (illustrated in the second column of Fig. \ref{plotMorph}), there is one dense spiral flow behind the companion and a second spiral emerging from a bow shock in front of the moving companion. This stable bow shock again shapes the global morphology into an approximate Archimedes spiral structure.
(iii) In the simulations in which $\varepsilon \gtrsim 3$ (illustrated in the right column of Fig. \ref{plotMorph}), an unstable bow shock forms in front of the companion, which results in a global morphology with irregular spiral structures.

In general more complex wind morphologies result in case the orbit is eccentric, as the phase-dependency makes the inner wind structure vary between the three types of inner wind structures described above throughout one orbital period. The details of how these different inner wind structures are formed and how they result in different global morphologies for both circular and eccentric configurations is described in detail by \citet{Malfait2021}. 

\subsection{Vertical wind extent \& distribution}
Next, the simulations can be categorised according to their three-dimensional density distribution. In the successors of AGB stars, being Post-AGB stars and PNe, circumbinary disks and bipolar outflows are observed, of which the formation mechanism is still uncertain \citep{vanWinckel2003,Bujarrabal2013,Oomen2020,Manick2021}. 
Studying the vertical extent of AGB winds may provide important information about the origin of these circumbinary disks and bipolar outflows.

There are two effects that can make the global wind distribution deviate from spherical symmetry.
Firstly, the orbital movement of the stars around the CoM induces a centrifugal force on the wind particles. This force gives the wind particles an additional acceleration in the orbital plane direction, which causes an elongation of the entire morphology, which we will refer to as flattening. Secondly, while the companion moves on its orbit, it gravitationally attracts matter. If this effect is strong, this can result in a density enhancement around the orbital plane, referred to as an equatorial density enhancement (EDE). We define an EDE to be present if the density around the orbital plane is strongly enhanced with the respect to a non-perturbed isotropic single star simulation, and if the density around the poles is decreased with respect to a single star simulation.
A more detailed explanation on how to determine if an EDE or flattening is present in the 3D simulations can be found in \cite{Malfait2021} and \cite{Maes2021}.

Table \ref{tableMorph}, adapted from \cite{Malfait2021}, illustrates that there is a flattening present in simulations where the gravitational pull of the companion is limited ($\varepsilon \lesssim 1$), and an EDE without flattening is found in more complex-structured simulations in which there is a strong gravitational impact of the companion ($\varepsilon \gtrsim 3$). 
Hence, this indicates that the impact of the orbital motion of the AGB star dominates when $\varepsilon \lesssim 1$, and the impact of the gravitational attraction of matter by the companion dominates in the simulations with high $\varepsilon$. Furthermore, it is important to note that an EDE and flattening occur for different simulation setups, and thereby a distinction should always be made.

\begin{table}[h!]
%	\centering
	\caption{Morphology classification.}
	{\begin{tabular}{@{\extracolsep{\fill}}clrllc}

		\midrule
		&&&&\\[-2ex]
		Model&   Global density distribution & & Meridional plane structure & Orbital Plane structure & $\varepsilon$\\
		&&&&\\[-2ex]
		\midrule
				
		v20e00  & Flattened              & no EDE  & Concentric arcs                   & Spiral - Archimedes & $1.0$ \\
		v20e25  & Flattened, asymmetric & no EDE  & Arcs                              & Spiral - Perturbed & $1.3$\\
		v20e50  & Flattened, asymmetric & no EDE  & Ring-arcs                         & Spiral - Perturbed & $2.0$\\
		v10e00  & Flattened           & with EDE & Bicentric rings - Peanut-shape             & Spiral - Archimedes & $2.6$ \\
		v10e25  & No flattening, irregular           & with EDE & irregular                             & irregular & $3.4$\\
		v10e50  & No flattening, irregular           & with EDE & irregular                             & irregular & $5.1$\\
		v05e00  & No flattening, irregular & with EDE & Rose                                          & Spiral - Squared & $4.1$ \\
		v05e25  & No flattening, irregular         & with EDE & Bipolar outflow                       & irregular & $5.5$\\
		v05e50  & No flattening, irregular          & with EDE & Bipolar outflow                       & irregular & $8.3$\\

		\midrule
	\end{tabular}}
%	\tabnote{\textit{Notes}: Morphology classification of the models, subdivided in: (i) global density distribution indicating if there is a flattening and/or EDE, (ii) structure in the meridional plane slice through the sink particles, (iii) wind structures in the orbital plane, and (iv) classification parameter $\varepsilon$ defined as in Eq.~(\ref{vareps}).}
	\label{tableMorph}
	{\footnotesize{\textbf{Notes.} 
			Morphology classification of the global density distribution, meridional plane and orbital plane structures, and value of classification parameter $\varepsilon$ (Eq. \ref{vareps}) of simulations with orbital separation $a = 6 \, \rm{au}$ and $M_{\rm{comp}} = 1 \rm{M_\odot}$. The model names give the values of the input wind velocity and eccentrictiy, with 'vXX' denoting the input wind velocity in $\rm{km \, s^{-1}}$ and 'eXX' the value of the eccentricity of the system multiplied by a factor 100. Table adapted from \citet{Malfait2021}.}}
\end{table}

\section{Impact of companion on terminal wind velocity}
\begin{figure}[h!]
	\centering
	\includegraphics[width = \textwidth]{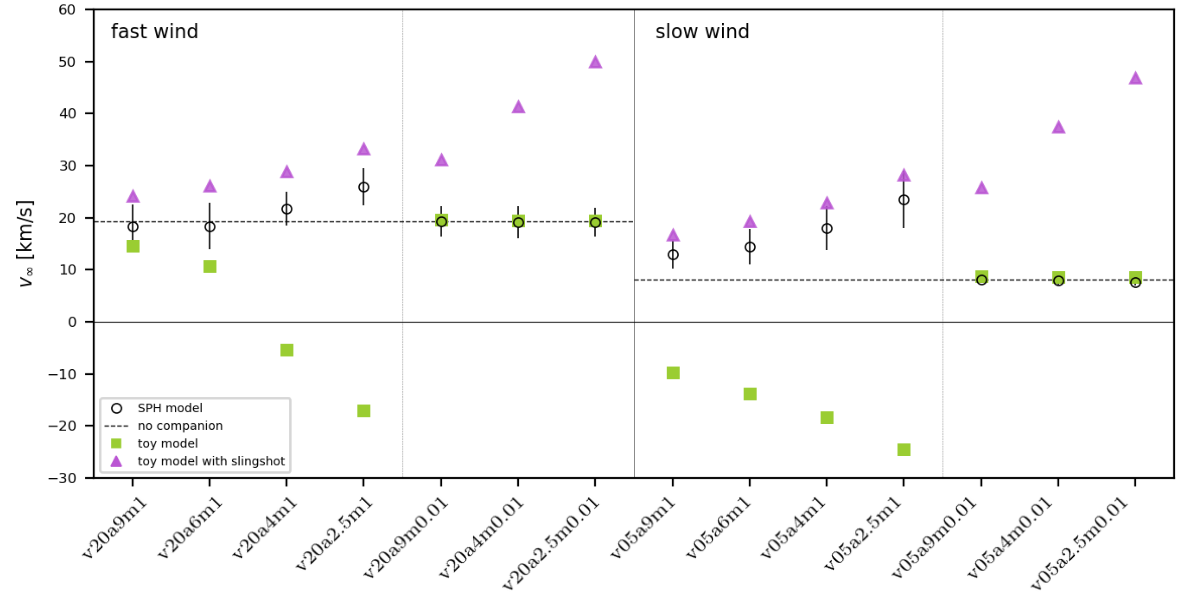}
	\caption{Terminal velocity for simulations with eccentricity $e=0$, as calculated from the toy model with and without a slingshot in triangles and squares, respectively, and from the simulation in empty circles. The terminal velocity of the corresponding single star model is given by the black dashed lines. The model names give the values of the input parameters, with 'vXX' denoting the input wind velocity in $\rm{km \, s^{-1}}$, 'aXX' the orbital separation in $\rm{au}$, and 'mXX' the companion mass in $\rm{M_\odot}$. Figure adapted from \citet{Maes2021}.}
	\label{velocityPlot}
\end{figure}

\citet{Maes2021} investigated the effect of a companion on the terminal expansion velocity of the wind, for simulations with a stellar or planetary companion, and with different initial wind velocity and orbital separation. Fig. \ref{velocityPlot} presents the results of their analysis. By comparing the terminal velocities of the simulations (indicated by empty dots) to the terminal velocities of a single star model (dashed line), it is clear that, whereas the impact of a planetary companion appears neglectable, a stellar companion does affect the expansion velocity of the wind. To investigate the cause of this deviation, a toy model was constructed in which the terminal velocity is calculated analytically, by only taking into account the effect of the gravitational potential of the companion. From the resulting toy model terminal velocities (squares in Fig. \ref{velocityPlot}) it can be concluded that for the case of a stellar companion, an important acceleration mechanism is missing, since too low and even negative terminal velocities result.
Therefore, the toy model is extended by including the gravitational slingshot mechanism, which states that by conservation of momentum and energy, a small object, moving past a larger body in motion, is accelerated or decelerated. With inclusion of the slingshot mechanism (triangles in Fig. \ref{velocityPlot}), the terminal velocity of the stellar models is a relatively good approximanion of the measured expansion velocity of the simulation. For more details see \citet{Maes2021}.

\section{Conclusion}
The impact of a binary companion on the outflow of an AGB star is studied using a grid of 3D hydrodynamic simulations constructed with the SPH code \textsc{Phantom}. From these simulations it is concluded that depending on the binary configuration, different inner wind structures and global morphologies result, varying from a regular Archimedes spiral with a spherically symmetric global density distribution, to highly perturbed spiral structures with equatorial density enhancements and flattened global morphologies. The classification parameter $\varepsilon$ is used to classify the morphology based on the characteristics of the binary system.  
Finally, we found that when the companion is massive enough, the terminal expansion velocity of the AGB wind is altered by the gravitational slingshot mechanism that acts on the wind particles.

%
%
%\begin{discussion}
%	
%\end{discussion}


\begin{thebibliography}{}

\bibitem[Bujarrabel \etal(2013)]{Bujarrabal2013}
{Bujarrabal, V., Alcolea, J., Van Winckel, H., \etal} 2013, \textit{A\&A}, 557, A104

\bibitem[Decin \etal(2020)]{Decin2020}
Decin, L., Montarg{\`e}s, M., Richards, A.~M.~S. \etal\ 2020, \textit{Science}, 369, 1497


\bibitem[El Mellah \etal(2020)]{ElMellah2020} 
El Mellah, I., Bolte, J., Decin, L. \etal\ 2020, \textit{A\&A}, 637, A91


\bibitem[H\"ofner \& Olofsson (2018)]{Hofner2018}
{H\"ofner, S., \& Olofsson, H.} 2018, \textit{A\&ARv}, 26, 1


\bibitem[Homan \etal (2020a)]{Homan2020a}
{Homan, W., Cannon, E., Montargès, M., \etal} 2020a, \textit{A\&A}, 642, A93

\bibitem[Homan \etal (2020b)]{Homan2020b}
{Homan, W., Montargès, M., Pimpanuwat, B., \etal} 2020b, \textit{A\&A}, 644, A61


\bibitem[Kervella \etal (2016)]{Kervella2016}
{Kervella, P., Homan, W., Richards, A. M. S., \etal} 2016, \textit{A\&A}, 596, A92

\bibitem[Kim \& Taam (2012)]{Kim2012}
{Kim, H., \& Taam, R. E.} 2012, \textit{ApJ}, 759, 59

	
\bibitem[Lamers \& Cassinelli (1999)]{Lamers1999}	
{Lamers, H. J. G. L. M., \& Cassinelli, J. P.} 1999, \textit{Introduction to Stellar Winds}

\bibitem[Maes \etal (2021)]{Maes2021}
{Maes, S., Homan, W., Malfait, J. \etal} 2021, \textit{A\&A}, 653, A25

\bibitem[Malfait \etal(2021)]{Malfait2021}
Malfait, J., Homan, W., Maes, S. \etal\ 2021, \textit{A\&A}, 652, A51

\bibitem[Manick \etal (2021)]{Manick2021}
{Manick, R., Miszalski, B., Kamath, D., \etal} 2021, \textit{MNRAS}, 508, 2226


\bibitem[Mastrodemos \& Morris (1998)]{Mastrodemos1998}
{Mastrodemos, N., \& Morris, M.} 1998, \textit{ApJ}, 497, 303
\bibitem[Mastrodemos \& Morris (1999)]{Mastrodemos1999}
{Mastrodemos, N., \& Morris, M. 1999}, \textit{ApJ}, 523, 357


\bibitem[Oomen \etal(2020)]{Oomen2020}
{Oomen, G.-M., Pols, O.,Van Winckel, H., Nelemans, G.} 2020, \textit{A\&A}, 642, A234

\bibitem[Price \etal(2018)]{Price2018}
{Price, D. J., Wurster, J., Tricco, T. S., \etal} 2018, \textit{PASA}, 35,
e031

\bibitem[Ramstedt \etal (2014)]{Ramstedt2014}
{Ramstedt, S., Mohamed, S., Vlemmings, W. H. T., \etal} 2014, \textit{A\&A}, 570, L14

\bibitem[Sahai \etal(2011)]{Sahai2011}
{Sahai, R., Morris, M. R., \& Villar, G. G.} 2011, \textit{AJ}, 141, 134
 
\bibitem[Theuns \& Jorissen (1993)]{Theuns1993} 
{Theuns, T., \& Jorissen, A.} 1993, \textit{MNRAS}, 265, 946
\bibitem[Theuns \etal (1996)]{Theuns 1996} 
{Theuns, T., Boffin, H. M. J., \& Jorissen, A.} 1996, \textit{MNRAS}, 280, 1264
 

\bibitem[Van Winckel (2003)]{vanWinckel2003}
{Van Winckel, H.} 2003, \textit{ARA\&A}, 41, 391



\end{thebibliography}
\end{document}